\documentclass[aoas,MSNbibl,nameyear,seceqn,dvips]{arximspdf}
\usepackage{multirow,dcolumn}
\usepackage{graphicx}

\doi{10.1214/14-AOAS746} 
\volume{8}
\issue{3}
\pubyear{2014}
\firstpage{1492}
\lastpage{1515}
\docsubty{FLA}

\makeatletter
\newcolumntype{d}[1]{D{.}{.}{#1}}
\renewcommand{\mid}{|}
\newcommand{\lleft}{\left}
\newcommand{\rright}{\right}
\newproclaim{remark}{Remark}[section]
\makeatother

\begin{document}
\begin{frontmatter}

\title{Two-phase sampling experiment for propensity score estimation
in self-selected samples}
\runtitle{Two-phase sampling experiment}

\begin{aug}
\author[A]{\fnms{Sixia}~\snm{Chen}\thanksref{T1}\ead[label=e1]{SixiaChen@westat.com}}
\and
\author[B]{\fnms{Jae-Kwang}~\snm{Kim}\corref{}\thanksref{T1,T2}\ead[label=e2]{jkim@iastate.edu}}
\runauthor{S. Chen and J.-K. Kim}
\affiliation{Westat and Iowa State University}
\address[A]{Westat\\
1600 Research Blvd\\
Rockville, Maryland 20850-3129\\
USA\\
\printead{e1}}
\address[B]{Department of Statistics\\
Iowa State University\\
Ames, Iowa 50011\\
USA\\
\printead{e2}} 
\end{aug}
\thankstext{T1}{Supported in part by a Cooperative Agreement
between the USDA Natural Resources Conservation Service and the Center for
Survey Statistics and Methodology at Iowa State University.}
\thankstext{T2}{Supported in part by NSF Grant MMS-121339.}

\received{\smonth{5} \syear{2013}}
\revised{\smonth{2} \syear{2014}}

%
\begin{abstract}
Self-selected samples are frequently obtained due to different levels
of survey participation propensity of the survey individuals. When the
survey participation is related to the survey topic of interest,
propensity score weighting adjustment using auxiliary information may
lead to biased estimation.
In this paper, we consider a parametric model for the response
probability that includes the study variable itself in the covariates
of the model and proposes
a novel application of two-phase sampling to estimate the parameters of
the propensity model. The proposed method includes an experiment in
which data are collected again from a subset of the original
self-selected sample.
With this two-phase sampling experiment, we can estimate the parameters
in a propensity score model consistently. Then the propensity score
adjustment can be applied to the self-selected sample to estimate the
population parameters.
Sensitivity of the selection model assumption is investigated from two
limited simulation studies.
The proposed method is applied to the 2012 Iowa Caucus Survey.
\end{abstract}

%
\end{frontmatter}

\section{Introduction}\label{sec1}

Nonresponse has become a major problem in sample surveys as
participation rates have declined in many surveys. When survey units
are chosen by surveyors, but these units elect not to participate,
self-selection occurs. If many units elect not to participate, the
representativeness of the observed sample can be called into question.
In this self-selected sample, the participation probabilities are
unknown and valid analysis of the self-selected sample is extremely
difficult when survey participation is related to survey items [\citet{r0}]. There exist sociological theories, such as the
leverage-saliency theory [\citet{r8}], that try to
identify psychological factors influencing survey participation, but it
is not clear how to use those theories to analyze observed data.

To reduce the bias of the estimator from self-selected samples, the
propensity score weighting method is commonly used. \citet{r21} as well as \citet{r22} proposed using propensity scores to
estimate treatment effects in observational studies.
\citet{r5} and \citet{r18} proposed nonresponse adjustment methods using regression
weighting techniques. \citet{r7} also considered a
nonlinear calibration procedure to control nonresponse bias.
\citet{r3} used the propensity score method to control
coverage bias in telephone surveys.
\citet{r15} applied the propensity score method to a volunteer panel web survey.
\citet{r4} used the method to address measurement error.
\citet{r16} and \citet{r24} considered the
propensity score method for a web-based voluntary sample. \citet{r14} provided some theory for the propensity score weighting
estimators.
All of these studies assumed an ignorable selection mechanism. That is,
it was assumed that the sample inclusion probability depends on one or
more auxiliary variables with known or estimated marginal
distributions. In other words, the selection mechanism was assumed to
be missing at random in the sense of \citet{r23}. If that is the case,
propensity scores can be consistently estimated and the resulting
analysis is valid under the assumed propensity score model.

In self-selected samples, the ignorable selection mechanism assumption
is not always realistic because survey participation may be related to
the survey topic of interest [\citet{r9}]. In
this case, the propensity model using only demographic auxiliary
variables may lead to biased estimation.
In this paper, we consider the nonignorable selection mechanism in the
propensity model for survey participation. To estimate the parameters
of the propensity model consistently, we propose a novel application of
the two-phase sampling experiment in a voluntary survey with voluntary
respondents contacted twice. Because the second-phase sample is
selected from those who already responded in the first phase, the model
parameters in the second-phase sampling mechanism are easy to estimate.
Furthermore, the estimated propensity for the second-phase sampling can
also provide useful information for the first-phase sampling mechanism
if the second contact is very similar to the original first contact.

Our paper is motivated by a telephone survey for the 2012 Iowa Caucus.
In this survey the individuals obtained from a probability sampling
procedure were asked about their intention to vote in the 2012 Iowa Caucus.
Low participation rate (15\%) may distort the representativeness of
the sample of respondents. In the beginning of the telephone survey,
the telephone interviewers identified the purpose of the study and
asked for participation in the survey. Thus, survey respondents who
chose to participate in this political survey may be systematically
different from those who refused to participate. Thus,
it is reasonable to assume that selection probability depends on the
study variables (intention to vote for given candidates). In November
2011, the first-phase self-selected sample was obtained and then the
second self-selected sample was obtained from the first-phase sample in
the next month. Due to the similarity of the survey questions for both
surveys, we treated the two selection mechanisms identically, up to
overall response rates. Then the model parameters were estimated from
the two-phase sample and the final estimates for voting intention were
computed using the estimated propensity. Further details are presented
in Section~\ref{sec6}.

\section{Basic setup}\label{sec2}

We introduce the basic setup which is very close to the 2012 Iowa
Caucus Survey (ICS). As presented in Figure~\ref{fig2} in Section~\ref{sec6}, the sampling
structure is a three-phase sampling, where the first-phase sample is a
probability sample but the second and the third samples are voluntary samples.

Let $U$ be a finite population of known size $N$, $A$ be a
probability-based sample with known first and second-order inclusion
probabilities, denoted by $\pi_{i}$ and $\pi_{ij}$, and $A_1 (\subset
A)$ be a respondent sample obtained from~$A$. In sample $A$, we observe~$x_{i}$, where $x_i$ is the vector of auxiliary variables which often
consist of demographic information. In sample $A_1$, we observe $(x'_i,
y_{1i})$, and $y_{1i}$ is the realized value of the study variable of
interest at the time of observing elements in $A_1$.
According to the leverage-saliency theory, the selection probability
for $A_1$ can be modeled as a function of $x_i$ and $z_i$, where $z_i$
is the unobservable variable that conceptually quantifies one's
interest on the survey topics. Such an assumption is reasonable if the
survey topic is informed to the survey interviewees in the very
beginning of the survey, which is the case with the 2012 Iowa Caucus Survey.
Thus, we may assume that
%
\begin{equation}
\operatorname{pr} ( \delta_{1i} =1 \mid x_{i}, y_{1i},
z_i ) = \frac{ \exp( \beta_0 + \beta_1' x_{i} + \beta_2 z_{i}) }{ 1+
\exp( \beta_0 + \beta_1' x_{i} + \beta_2 z_{i})} \label{0}
\end{equation}
for some coefficient $(\beta_0, \beta_1, \beta_2)$, where $\delta
_{1i}$ is the indicator function for element $i$ to be in sample $A_{1}$.

Now, as $z_i$ is not observable, we consider an alternative model
using an observed surrogate variable for $z$. Because we observe $y_1$
instead of $z$, a natural alternative model is
%
\begin{equation}
\pi_{1i}(\phi)=\operatorname{pr} ( \delta_{1i} =1 \mid
x_{i}, y_{1i} ) = \frac{ \exp( \phi_0 + \phi_1' x_{i} + \phi_2 y_{1i}) }{ 1+ \exp
( \phi_0 + \phi_1' x_{i} + \phi_2 y_{1i})}, \label{1}
\end{equation}
for some $(\phi_0, \phi_1, \phi_2)$.
More generally, we can write $\pi_{1i} (\phi) = \pi(x_i, y_{1i};
\phi)$ for known function $\pi( \cdot)$ with unknown parameter $\phi
=(\phi_0, \phi_1, \phi_2)$.

To estimate the parameters in (\ref{1}), we subject the respondents of
the first-phase sampling to similar survey questions and obtain a
second respondent sample $A_2$ from $A_1$.
That is, we perform a two-phase sampling under the same response mechanism.
The response model for $A_2$ is
%
\begin{equation}
\qquad \pi_{2i}\bigl(\phi^*\bigr)=\operatorname{pr} ( \delta_{2i} =1 \mid
x_{i}, y_{2i}, \delta_{1i}=1 ) =
\frac{\exp( {\phi}_0^* + \phi_1'
x_{i} + \phi_2 y_{2i}) }{ 1+ \exp( {\phi}_0^* + \phi_1' x_{i} +
\phi_2 y_{2i}) }, \label{2}
\end{equation}
where $\delta_{2i}$ is the indicator function for element $i$ to be in
sample $A_{2}$, $y_{2i}$ is the measurement of $Y$ at the time of
selecting $A_2$, and $(\phi_1', \phi_2)$ is defined in (\ref{1}).
Here, we allow the study item $Y$ to be time-dependent; in other words,
the value of $Y$ can change over time.
Thus, we assume that the conditional odds for the first-respondent
selection and for the second-respondent selection are the same.
Obtaining the second-respondent sample $A_2$ from $A_1$ is an
experiment that we perform to understand
the response mechanism of $A_1$ from $A$ by asking the same\vadjust{\goodbreak} questions
to the same people whose $y$-values are available.
We may perform the experiment in the random subsample of $A_1$ in order
to reduce the cost, but the subsampling does not make any difference in
the resulting analysis.
From the two-respondent sample, we are interested in estimating $\theta
_1=E(Y_1)$ and $\theta_2=E(Y_2)$.

We now discuss parameter estimation for the propensity models. Note
that we observe $(x_{i}', y_{1i}, y_{2i})$ in $A_2$. Thus, we can
construct the following estimating equation to estimate the parameters
in (\ref{2}):
%
\begin{equation}
\sum_{i \in A_{1}}\omega_{i} \biggl\{
\frac{\delta_{2i}}{ \pi
_{2i}(\phi^{*})} -1 \biggr\}h_{1i}=0, \label{3}
\end{equation}
where $\omega_{i}=\pi^{-1}_{i}$, $h_{1i} = (1, x'_{i}, y_{1i})'$.
Once $\hat{\phi}^*=(\hat{\phi}^{*}_0, \hat{\phi}_1', \hat{\phi
}_2)'$ is computed, we can use
\[
\sum_{i \in A_1} \frac{\omega_{i}}{ \pi_{1i}( \phi_0, \hat{\phi
}_1, \hat{\phi}_2 )} = N
\]
to estimate $\phi_0$. Equation (\ref{3}) is a calibration equation in
the second-phase respondent sample using $h_{1i}$ as the control
variable. Use of calibration for propensity score adjustment has been
considered by \citet{r5}, \citet{r11} and \citet{r12}.

Once the parameters in (\ref{1}) and (\ref{2}) are estimated, we can
use the following propensity-score-adjusted estimator:
%
\begin{equation}
\hat{\theta}_1 =\frac{1}{N} \sum
_{i \in A_1}\omega_{i} \hat{\pi }_{1i}^{-1}
y_{1i} \label{1-5}
\end{equation}
to estimate $\theta_1=E(Y_1)$. Also, we can use
%
\begin{equation}
\hat{\theta}_2 = \frac{1}{N}\sum_{i \in A_2}
\omega_{i}\hat{\pi }_{1i}^{-1} \hat{
\pi}_{2i}^{-1} y_{2i} \label{1-6}
\end{equation}
to estimate $\theta_2=E(Y_2)$.
In addition, we may use the population-level information of $x$ to
improve the efficiency of the propensity-score-adjusted estimators,
which will be presented in Section~\ref{sec4}.

\section{Main results}\label{sec3}

In this section we discuss some asymptotic properties of the proposed
propensity-score-adjusted estimators.
To discuss asymptotic\vspace*{2pt} properties of $\hat{\theta}_1$ in (\ref{1-5}),
we first define $\Phi=(\phi^{*}_{0},\phi_{1}',\phi_{2},\phi_0)'$,
%
\begin{equation}
U_1 (\Phi) \equiv\sum_{i \in A_1}
\omega_{i} \biggl\{ \frac{\delta
_{2i}}{\pi_{2i} (\phi^{*}_{0}, \phi_{1},\phi_{2} )} -1 \biggr\} \bigl(1,
x'_{i}, y_{1i} \bigr)' =
(0,0,0)' \label{2-1}
\end{equation}
and
%
\begin{equation}
U_2 (\Phi) \equiv\sum_{i \in A_1}
\frac{\omega_{i}}{\pi_{1i}
(\phi_{0}, \phi_{1},\phi_{2})} - N = 0. \label{2-2}
\end{equation}
Thus, equations (\ref{2-1}) and (\ref{2-2}) are a system of nonlinear
equations that can be solved for $\Phi$. We can write $U_c(\Phi)' =
 [ U_1 (\Phi)', U_2 (\Phi)'  ]$, and $(\hat{\theta}_1,
\widehat{\Phi}')'$ can be\vadjust{\goodbreak} obtained as the solution to
\begin{eqnarray*}
U_p (\theta_1, \Phi) &=& 0,\qquad
U_c (\Phi) = 0,
\end{eqnarray*}
where $U_p (\theta_1, \Phi) = N^{-1}\sum_{i \in A_1}\omega_{i}
 \{ \pi_{1i} (\phi_{0}, \phi_{1},\phi_{2})  \}^{-1}
y_{1i} - \theta_1 $.
Because
\[
E \bigl\{ U_p \bigl(\theta_1^*, \Phi^*\bigr)  \bigr\} = 0\quad\mbox{and}\quad E \bigl\{ U_c \bigl(\theta_1^*, \Phi^*\bigr)  \bigr\} = 0,
\]
where $(\theta _1^*, \Phi^{* '} )'$ is the true parameter value, the solution $(\hat
{\theta}_1, \widehat{\Phi}')'$ is consistent and has asymptotic variance
\begin{eqnarray}
\qquad\operatorname{var} \pmatrix{ \hat{\theta}_1
\vspace*{3pt}\cr
\widehat{\Phi}} &\cong&
\lleft\{ \matrix{ -1 & E ( \partial U_p/ \partial\Phi )
\vspace*{3pt}\cr
0 &
E ( \partial U_c/ \partial\Phi ) } %
 \rright
\}^{-1} \lleft\{ %
\matrix{ \operatorname{var} (
U_p ) & \operatorname{cov} ( U_p, U_c )
\vspace*{3pt}\cr
\operatorname{cov} ( U_c, U_p ) & \operatorname{var}
( U_c )} %
 \rright\}
\nonumber
\nonumber\\[-8pt]\\[-8pt]\nonumber
&&{}\times \lleft[ \lleft\{ %
\matrix{ -1 & E ( \partial
U_p/ \partial\Phi )
\vspace*{3pt}\cr
0 & E ( \partial U_c/ \partial
\Phi ) } %
 \rright\}' \rright]^{-1}.
\end{eqnarray}
Using
\[
\lleft[ %
\matrix{ -1 & E ( \partial U_p/ \partial\Phi
)
\vspace*{3pt}\cr
0 & E ( \partial U_c/ \partial\Phi ) } %
 \rright]^{-1} = \lleft[ %
\matrix{ -1 & E ( \partial
U_p/ \partial\Phi ) \bigl\{E(\partial U_{c}/\partial\Phi)
\bigr\}^{-1}
\vspace*{3pt}\cr
0 & \bigl\{E ( \partial U_c/ \partial\Phi
) \bigr\}^{-1}} %
 \rright], %
\]
then the asymptotic variance of $\hat{\theta}_{1}$ can be written,
using the definition of $U_{p}$~and~$U_c$, as
\begin{eqnarray}
\operatorname{var} ( \hat{\theta}_1 ) &\cong& \operatorname{var}
\biggl\{ U_{p}-E\biggl(\frac{\partial U_{p}}{\partial\Phi
}\biggr) \biggl\{E\biggl(
\frac{\partial U_{c}}{\partial\Phi}\biggr) \biggr\} ^{-1}U_{c} \biggr\}
\nonumber
\\
&=& \operatorname{var}\lleft\{ \hat{\theta}_{1}(\Phi)-E \biggl
\{ \frac{
\partial}{ \partial\Phi} \hat{\theta}_1 (\Phi) \biggr\}\lleft[
\matrix{ E\bigl\{ \partial U_1 (\Phi) /\partial\Phi\bigr
\}
\vspace*{3pt}\cr
E\bigl\{ \partial U_2 (\Phi)/ \partial\Phi\bigr\} } %
 \rright]^{-1}U_{c} \rright\},
\nonumber
\end{eqnarray}
where $\hat{\theta}_1 (\Phi) = N^{-1} \sum_{i \in A_1}\omega_{i}
y_{1i} \{ 1+ \exp ( - \phi_0 - \phi'_1 x_{i} - \phi_2 y_{1i}
 ) \}$.
Thus, the asymptotic variance can be written as
%
\begin{eqnarray}\label{2-3}
\operatorname{var} ( \hat{\theta}_1 ) &\cong& \frac{1}{N^2}V
\Biggl[\sum_{i=1}^N \omega_{i}
\delta_{i}\frac{\delta_{1i}}{\pi
_{1i}} y_{1i} - B_{1,y} \Biggl
\{ \sum_{i=1}^N \omega_{i}\delta
_{i}\delta_{1i} \biggl( \frac{\delta_{2i}}{\pi_{2i}} - 1
\biggr)h_{1i} \Biggr\}
\nonumber\\[-8pt]\\[-8pt]\nonumber
&&\hspace*{136pt}{}- B_{2,y}\sum_{i=1}^N \biggl(
\omega_{i}\delta_{i}\frac{\delta
_{1i}}{\pi_{1i}} -1 \biggr) \Biggr],
\end{eqnarray}
and
\begin{eqnarray*}
(B_{1,y}, B_{2,y}) &=& N \times E \biggl\{ \frac{ \partial}{ \partial\Phi}
\hat{\theta }_1 (\Phi) \biggr\}\lleft[ %
\matrix{ E
\bigl\{ \partial U_1 (\Phi) /\partial\Phi\bigr\}
\vspace*{3pt}\cr
E\bigl\{
\partial U_2 (\Phi)/ \partial\Phi\bigr\} } %
 \rright]^{-1}
\\
&=& \sum_{i=1}^{N}(1-\pi_{1i})y_{1i}
\bigl(0,x'_{i},y_{1i},1\bigr)\pmatrix{
\displaystyle\sum_{i=1}^{N}\pi_{1i}(1-\pi_{2i})h_{1i}h'_{2i},0_{r \times1}
\vspace*{3pt}\cr
\displaystyle\sum_{i=1}^{N}(1-\pi_{1i}) \bigl(0,x'_{i},y_{1i},1\bigr)}^{-1},
\end{eqnarray*}
where $h_{2i}=(1,x'_{i},y_{2i})'$ and $0_{r \times1}$ is the vector of
zeros with dimension $r \times1$, with $r=2+p$, and $p$ is the
dimension of $x_{i}$.
Note that the variance (\ref{2-3}) can be written~as
%
\begin{eqnarray}\label{2-4-1}
\operatorname{var} ( \hat{\theta}_1 ) &=& \operatorname{var} \Biggl
\{ \frac{1}{N}\sum^{N}_{i=1}
\omega_{i}\delta_{i}\frac{\delta
_{1i}}{\pi_{1i}}(y_{1i}-B_{2,y})
\Biggr\}
\nonumber\\[-8pt]\\[-8pt]\nonumber
&&{}+ \operatorname{var} \Biggl\{ \frac{1}{N}\sum
^{N}_{i=1}B_{1,y}\omega _{i}
\delta_{i}\delta_{1i}\biggl(\frac{\delta_{2i}}{\pi
_{2i}}-1
\biggr)h_{1i} \Biggr\}.
\end{eqnarray}
Roughly speaking, the first term in (\ref{2-4-1}) is the asymptotic
variance of the propensity-score-adjusted estimator when $(\phi'_1,
\phi_2)$ is known and the second term is the additional variance due
to the fact that $(\phi'_1, \phi_2)$ is estimated from the
second-phase sample.
Variance estimation is straightforward from (\ref{2-3}), as we can
replace the unknown parameters with their estimators.

We now discuss the asymptotic properties of $\hat{\theta}_2$ in (\ref
{1-6}), that is, the direct propensity-score-adjusted estimator of
$\theta_2$.
Using the argument similar to (\ref{2-3}), we can obtain
%
\begin{eqnarray}\label{2-5}
\qquad\operatorname{var} ( \hat{\theta}_2 ) &\cong& \frac{1}{N^2}V
\Biggl[\sum_{i=1}^N \omega_{i}
\delta_{i}\frac{\delta_{1i}\delta
_{2i}}{\pi_{1i} \pi_{2i}} y_{2i} - D_{1,y} \Biggl
\{ \sum_{i=1}^N \omega_{i}
\delta_{i}\delta_{1i} \biggl( \frac{\delta_{2i}}{\pi
_{2i}} - 1
\biggr)h_{1i} \Biggr\}
\nonumber\\[-8pt]\\[-8pt]\nonumber
&&\hspace*{150pt}{}- D_{2,y}\sum_{i=1}^N \biggl(
\omega_{i}\delta_{i}\frac{\delta
_{1i}}{\pi_{1i}} -1 \biggr) \Biggr],
\end{eqnarray}
where
\begin{eqnarray*}
(D_{1,y}, D_{2,y}) &=& N \times E \biggl\{ \frac{ \partial}{ \partial\Phi}
\hat{\theta }_2 (\Phi) \biggr\}\lleft[ %
\matrix{ E\bigl
\{ \partial U_1 (\Phi) /\partial\Phi\bigr\}
\vspace*{3pt}\cr
E\bigl\{ \partial
U_2 (\Phi)/ \partial\Phi\bigr\} } %
 \rright]^{-1}
\\
&=& \sum^{N}_{i=1}y_{2i} \bigl\{
(1-\pi _{1i}) \bigl(0,x'_{i},y_{1i},1
\bigr)+(1-\pi_{2i}) \bigl(1,x'_{i},y_{2i},0
\bigr) \bigr\}
\nonumber
\\
&&{}\times \pmatrix{
\displaystyle\sum_{i=1}^{N}\pi_{1i}(1-\pi_{2i})h_{1i}h'_{2i},0_{r \times1}
\vspace*{3pt}\cr
\displaystyle\sum_{i=1}^{N}(1-\pi_{1i}) \bigl(0,x'_{i},y_{1i},1\bigr)}^{-1}
\nonumber
\end{eqnarray*}
and we used
\begin{eqnarray*}
\hat{\theta}_2 (\Phi) &=& \frac{1}{N}\sum
_{i \in A_{2}}\omega _{i}y_{2i} \bigl\{ 1+\exp
\bigl(-\phi_{0}-\phi'_{1}x_{i}-\phi
_{2}y_{1i}\bigr) \bigr\}
\\
&&{}\times \bigl\{ 1+\exp\bigl(-
\phi^{*}_{0}-\phi '_{1}x_{i}-
\phi_{2}y_{2i}\bigr) \bigr\}.
\end{eqnarray*}
Writing
\[
\frac{\delta_{1i}\delta_{2i}}{\pi_{1i} \pi_{2i}} y_{2i} = \frac{\delta_{1i}}{\pi_{1i} } y_{2i} +
\frac{\delta_{1i}}{\pi
_{1i}} \biggl( \frac{\delta_{2i}}{\pi_{2i}} - 1 \biggr) y_{2i},
\]
the asymptotic variance in (\ref{2-5}) can be written as
%
\begin{eqnarray}
\label{2-5b} \operatorname{var} ( \hat{\theta}_2 )&\cong&
\frac{1}{N^{2}}\operatorname{var} \Biggl[ \sum_{i=1}^{N}
\omega_{i}\delta _{i}\frac{\delta_{1i}}{\pi_{1i}}(y_{2i}-D_{2,y})
\Biggr]
\nonumber\\[-8pt]\\[-8pt]\nonumber
&&{}+ \frac{1}{N^{2}}\operatorname{var} \Biggl\{ \sum
_{i=1}^N \omega _{i}\delta_{i}
\frac{ \delta_{1i}}{\pi_{1i}} \biggl( \frac{ \delta
_{2i}}{\pi_{2i}} - 1 \biggr) (y_{2i} -
D_{1,y} \pi_{1i}h_{1i}) \Biggr\}.
\end{eqnarray}
Thus, comparing (\ref{2-5b}) with (\ref{2-5}), we note that $\hat
{\theta}_2$ based on the $A_2$ sample does not necessarily have a
larger variance than $\hat{\theta}_1$ that is computed from the
first-phase sample $A_1$. In fact, if $y_{2i}$ is well approximated by
$h_{1i}$, then the second term of the asymptotic variance in (\ref
{2-5b}) is small and $\hat{\theta}_2$ is more efficient than $\hat
{\theta}_1$.

Instead of using the direct estimator $\hat{\theta}_{2}$ in (\ref
{1-6}), we can use a two-phase regression estimator to improve
efficiency. The estimator is efficient in that it incorporates
auxiliary information obtained from the first-phase sampling. See
\citet{r10}, \citet{r17} and \citet{r13} for more details about two-phase regression estimators. In
our setup, the data vector $h_{1i}= (1, x'_{i}, y_{1i})'$ is available
for both $A_1$ and $A_2$. Thus, the two natural estimators for the
population mean $\bar{h}_{1N}=N^{-1}\sum^{N}_{i=1}h_{1i}$, $\hat
{h}_{1,1}=N^{-1}\sum_{i \in A_{1}}\omega_{i}\hat{\pi
}^{-1}_{1i}h_{1i}$ and $\hat{h}_{2,1}=N^{-1}\sum_{i \in A_{2}}\omega
_{i}\hat{\pi}^{-1}_{1i} \hat{\pi}_{2i}^{-1} h_{1i}$ can be computed
from $A_1$ and $A_2$, respectively, and they are both approximately
unbiased for $\bar{h}_{1N}$. Using $\hat{h}_{1,1} $ and $\hat{h}_{2,1}$,
the two-phase regression estimator can be constructed by
%
\begin{eqnarray}
\hat{\theta}_{2,\mathrm{Reg}} &=& \hat{\theta}_{2}-\widehat{C}_{h_{1}}
(\hat {h}_{2,1}-\hat{h}_{1,1} ), 
\label{reg}
\end{eqnarray}
where
%
\begin{equation}
\widehat{C}_{h_{1}}=\sum_{i \in A_{2}}
\omega_{i}\hat{\pi }^{-1}_{1i}\hat{
\pi}^{-1}_{2i}y_{2i}h'_{1i}
\biggl\{ \sum_{i \in
A_{2}}\omega_{i}\hat{
\pi}^{-1}_{1i}\hat{\pi }^{-1}_{2i}h_{1i}h'_{1i}
\biggr\}^{-1}. \label{reg2}
\end{equation}
%
Because $E(\hat{h}_{2,1}-\hat{h}_{1,1}) \cong0$, the regression
estimator in (\ref{reg}) is approximately unbiased, regardless of the
choice of $\widehat{C}_{h_{1}}$. By applying the linearization method to
each term of (\ref{reg}), we can get
\begin{eqnarray*}
\hat{\theta}_{2,\mathrm{Reg}} &\cong& \frac{1}{N} \sum
_{i=1}^N \biggl\{ \omega _{i}
\delta_{i}\frac{\delta_{1i}\delta_{2i}}{\pi_{1i} \pi_{2i}} y_{2i} - D_{1i,\mathrm{Reg}}
\omega_{i}\delta_{i}\delta_{1i} \biggl(
\frac
{\delta_{2i}}{\pi_{2i}} - 1 \biggr)h_{1i}
\\
&&\hspace*{126pt}{} - D_{2,\mathrm{Reg}} \biggl( \omega
_{i}\delta_{i}\frac{\delta_{1i}}{\pi_{1i}} -1 \biggr) \biggr\},
\end{eqnarray*}
where $D_{1i,\mathrm{Reg}}=D_{1,y}+C^{*}_{h_{1}}\pi
^{-1}_{1i}-C^{*}_{h_{1}}(D_{1,h_{1}}-B_{1,h_{1}})$ and
$D_{2,\mathrm{Reg}}=D_{2,y}-C^{*}_{h_{1}}(D_{2,h_{1}}-B_{2,h_{1}})$ with
\begin{eqnarray*}
(B_{1,h_{1}},B_{2,h_{1}})&=&\sum_{i =1}^{N}(1-
\pi _{1i})h_{1i}\bigl(0,x'_{i},y_{1i},1
\bigr)
\\
&&{}\times \pmatrix{
\displaystyle\sum_{i =1}^{N}\pi_{1i}(1-\pi_{2i})h_{1i}h'_{2i},0_{r \times1}
\vspace*{3pt}\cr
\displaystyle\sum_{i =1}^{N}(1-\pi_{1i}) \bigl(0,x'_{i},y_{1i},1
\bigr)}^{-1}, 
\end{eqnarray*}
$C_{h_{1}}^*= p\lim\widehat{C}_{h_{1}}$,
\begin{eqnarray*}
(D_{1,h_{1}}, D_{2,h_{1}}) &=& \sum_{i =1}^{N}h_{1i}
\bigl\{ (1-\pi_{1i}) \bigl(0,x'_{i},y_{1i},1
\bigr)+(1-\pi _{2i}) \bigl(1,x'_{i},y_{2i},0
\bigr) \bigr\}
\\
&&{}\times \pmatrix{
\displaystyle\sum_{i =1}^{N}
\pi_{1i}(1-\pi_{2i})h_{1i}h'_{2i},
0_{r \times1}
\vspace*{3pt}\cr
\displaystyle\sum_{i =1}^{N}(1-
\pi_{1i}) \bigl(0,x'_{i},y_{1i},1
\bigr)}^{-1}.
\nonumber
\end{eqnarray*}
Thus, the asymptotic variance is
\begin{eqnarray}
\operatorname{var} ( \hat{\theta}_{2,\mathrm{Reg}} ) &\cong& \frac
{1}{N^2}
\operatorname{var} \Biggl[\sum_{i=1}^N
\omega_{i}\delta_{i} \frac
{\delta_{1i}\delta_{2i}}{\pi_{1i} \pi_{2i}} y_{2i} -
\Biggl\{ \sum_{i=1}^N \omega_{i}
\delta_{i}D_{1i,\mathrm{Reg}}\delta_{1i} \biggl(
\frac
{\delta_{2i}}{\pi_{2i}} - 1 \biggr)h_{1i} \Biggr\}
\nonumber
\\
&&\hspace*{160pt}{}- D_{2,\mathrm{Reg}}\sum_{i=1}^N \biggl(
\omega_{i}\delta_{i}\frac{\delta
_{1i}}{\pi_{1i}} -1 \biggr) \Biggr]
\nonumber
\\
&=& \frac{1}{N^{2}}\operatorname{var} \Biggl[ \sum
_{i=1}^{N}\omega _{i}\delta_{i}
\frac{\delta_{1i}}{\pi_{1i}}(y_{2i}-D_{2,\mathrm{Reg}}) \Biggr]
\nonumber
\\
&&{}+ \frac{1}{N^{2}}\operatorname{var} \Biggl[ \sum
_{i=1}^{N}\omega _{i}\delta_{i}
\frac{\delta_{1i}}{\pi_{1i}}\biggl(\frac{\delta_{2i}}{\pi
_{2i}}-1\biggr) (y_{2i}-D_{1i,\mathrm{Reg}}
\pi_{1i}h_{1i}) \Biggr].
\nonumber
\end{eqnarray}

%
\begin{remark}\label{rem3.1}
Instead of $\widehat{C}_{h_{1}}$ in (\ref{reg2}), the optimal choice of
$\widehat{C}_{h_{1}}$ that minimizes the variance among the class of
regression estimators with a form specified by (\ref{reg}) is
%
\begin{equation}
\widehat{C}_{h_{1},\mathrm{opt}} =\widehat{\operatorname{cov}} ( \hat{\theta }_{2},
\hat{h}_{2,1}-\hat{h}_{1,1} ) \bigl\{\widehat{\operatorname{var}}(
\hat{h}_{2,1}-\hat{h}_{1,1} ) \bigr\}^{-1},
\label{Copt}
\end{equation}
where
\begin{eqnarray*}
\widehat{\operatorname{cov}} ( \hat{\theta}_{2},\hat{h}_{2,1}-
\hat {h}_{1,1} ) &\cong& \frac{1}{N^{2}}\sum
_{i \in A_{2}}\omega _{i}\frac{1-\hat{\pi}_{2i}}{\hat{\pi}^{2}_{2i}}\biggl(
\frac
{y_{2i}}{\hat{\pi}_{1i}}-\widehat{D}_{1,y}h_{1i}\biggr)\hat{\eta}'_{i}
\\
&&{}-\frac
{1}{N^{2}}\sum
_{i \in A_{2}}\omega_{i}\frac{1-\hat{\pi}_{1i}}{\hat{\pi}^{2}_{1i}}\biggl(
\frac{y_{2i}}{\hat{\pi}_{2i}}-\frac{\widehat{D}_{2,y}}{\hat{\pi}_{2i}}\biggr)\hat{\tau}'
\nonumber
\\
&&{}+ \frac{1}{N^{2}}\sum_{i \in A}\sum
_{j \in A}\widetilde{\Delta }_{ij}\frac{\hat{z}_{3i}}{\pi_{i}}
\frac{\hat{z}^{\prime}_{4j}}{\pi
_{j}},
\\
\widehat{\operatorname{var}}( \hat{h}_{2,1}-\hat{h}_{1,1} ) &
\cong& \frac
{1}{N^{2}}\sum_{i \in A_{2}}
\omega_{i}\frac{1-\hat{\pi}_{2i}}{\hat
{\pi}^{2}_{2i}}\hat{\eta}^{\otimes2}_{i}
\\
&&{}+
\frac{1}{N^{2}}\sum_{i
\in A_{1}}\omega_{i}
\frac{1-\hat{\pi}_{1i}}{\hat{\pi
}^{2}_{1i}}\hat{\tau}^{\otimes2}
\nonumber
\\
&&{}+ \frac{1}{N^{2}}\sum_{i \in A}\sum
_{j \in A}\widetilde{\Delta }_{ij}\frac{\hat{z}_{4i}}{\pi_{i}}
\frac{\hat{z}^{\prime}_{4j}}{\pi
_{j}},
\end{eqnarray*}
where $M^{\otimes2}=MM^{\prime}$,
\begin{eqnarray*}
\hat{z}_{3i}&=&\frac{\delta_{1i}}{\hat{\pi}_{1i}} \biggl\{\frac
{\delta_{2i}}{\hat{\pi}_{2i}}y_{2i}-
\biggl(\frac{\delta_{2i}}{\hat{\pi
}_{2i}}-1\biggr)\hat{\pi}_{1i}\widehat{D}_{1,y}h_{1i}-
\widehat{D}_{2,y} \biggr\},
\\
\hat{z}_{4i}&=&\frac{\delta_{1i}}{\hat{\pi}_{1i}} \biggl\{ \biggl(\frac
{\delta_{2i}}{\hat{\pi}_{2i}}-1
\biggr) \bigl(\hat{\pi}^{-1}_{1i}-\widehat{D}_{1,h_{1}}+
\widehat{B}_{1,h_{1}}\bigr)\hat{\pi}_{1i}h_{1i}-(\widehat{D}_{2,h_{1}}-\widehat{B}_{2,h_{1}}) \biggr\}
\end{eqnarray*}
%
and
$\hat{\eta}_{i}= h'_{1i}\hat{\pi}^{-1}_{1i}-(\widehat{D}_{1,h_{1}}-\widehat{B}_{1,h_{1}})h'_{1i}$, $\hat{\tau}=\widehat{D}_{2,h_{1}}-\widehat{B}_{2,h_{1}}$ with
\begin{eqnarray*}
(\widehat{B}_{1,h_{1}},\widehat{B}_{2,h_{1}})&=&\sum
_{i \in A_{1}}\omega _{i}\frac{1-\hat{\pi}_{1i}}{\hat{\pi
}_{1i}}h_{1i}
\bigl(0,x'_{i},y_{1i},1\bigr)
\\
&&{}\times \pmatrix{
\displaystyle\sum_{i \in A_{2}}\omega_{i}\hat{\pi}^{-1}_{2i}(1-\hat{\pi }_{2i})h_{1i}h'_{2i},0_{r \times1}
\vspace*{3pt}\cr
\displaystyle\sum_{i \in A_{1}}\omega_{i}\hat{\pi}^{-1}_{1i}(1-\hat{\pi }_{1i})\bigl(0,x'_{i},y_{1i},1\bigr)}^{-1},
\label{eB1h1B2h1}
\\
(\widehat{D}_{1,h_{1}}, \widehat{D}_{2,h_{1}}) &=& \sum
_{i \in A_{2}}\omega_{i}\frac{h_{1i}}{\hat{\pi}_{1i}\hat{\pi
}_{2i}} \bigl\{ (1-\hat{
\pi}_{1i}) \bigl(0,x'_{i},y_{1i},1
\bigr)+(1-\hat{\pi }_{2i}) \bigl(1,x'_{i},y_{2i},0
\bigr) \bigr\}
\nonumber
\\
&&{}\times \pmatrix{
\displaystyle \sum_{i \in A_{2}}\omega_{i}\hat{\pi}^{-1}_{2i}(1-\hat{\pi }_{2i})h_{1i}h'_{2i},0_{r \times1}
\vspace*{3pt}\cr
\displaystyle\sum_{i \in A_{1}}\omega_{i}\hat{\pi}^{-1}_{1i}(1-\hat{\pi }_{1i})\bigl(0,x'_{i},y_{1i},1\bigr)}^{-1}
\label{eD1h1D2h1}
\end{eqnarray*}
and
\begin{eqnarray*}
(\widehat{D}_{1,y}, \widehat{D}_{2,y}) &=& \sum
_{i \in A_{2}}\omega_{i}\frac{y_{2i}}{\hat{\pi}_{1i}\hat
{\pi}_{2i}} \bigl\{ (1-\hat{
\pi}_{1i}) \bigl(0,x'_{i},y_{1i},1
\bigr)+(1-\hat {\pi}_{2i}) \bigl(1,x'_{i},y_{2i},0
\bigr) \bigr\}
\nonumber
\\
&&{}\times \pmatrix{
\displaystyle\sum_{i \in A_{2}}\omega_{i}\hat{\pi}^{-1}_{2i}(1-\hat{\pi }_{2i})h_{1i}h'_{2i},0_{r \times1}
\vspace*{3pt}\cr
\displaystyle\sum_{i \in A_{1}}\omega_{i}\hat{\pi}^{-1}_{1i}(1-\hat{\pi }_{1i})\bigl(0,x'_{i},y_{1i},1\bigr)}^{-1}.
\end{eqnarray*}
Such an estimator can be called a design-optimal regression estimator,
as termed by \citet{r20}.
\end{remark}
%

%
\begin{remark}\label{rem3.2}
Model (\ref{0}) can be viewed as a measurement error model in the
sense that the true covariate $z_i$ is not observed, but surrogate
variables $y_{1i}=z_i+ u_{1i}$ and $y_{2i}=z_i + u_{2i}$ are observed
instead of $z_i$, where $u_{1i}$ and $u_{2i}$ are measurement errors
associated with $y_{1i}$ and $y_{2i}$, respectively. Assuming that the
measurement errors follow a normal distribution, we can apply the
propensity weighting method with error-prone covariates, using the
recent approach of \citet{r19}, to our
two-phase sampling experiment.

Specifically, the measurement error model for obtaining the $A_1$
sample is given by (\ref{0}) with $y_{1i} \sim N(z_i, \sigma_{u1}^2)$
and the measurement error model for obtaining the~$A_2$ sample from
$A_1$ sample is given by
\[
\operatorname{pr} ( \delta_{2i} =1 \mid x_i, y_{2i},
z_i, \delta _{1i}=1 ) = \frac{ \exp(\beta_0^* + \beta_1' x_i + \beta_2 z_i) }{ 1+ \exp
(\beta_0^* + \beta_1' x_i + \beta_2 z_i) },
\]
with $y_{2i} \sim N(z_i, \sigma_{u2}^2)$. It turns out that our
proposed method is equivalent to the propensity weighting method under
this measurement error model using the method of \citet{r19}. More details can be found in Section~B of the supplemental
article [\citeauthor{r1b} (\citeyear{r1a,r1b})].
\end{remark}

\section{Use of population auxiliary information}\label{sec4}

In this section we assume that population information $\overline{X}_{N}=\sum_{i \in U} x_i$ is available. In this case, we can apply
the calibration techniques [\citet{r2}; \citet{r6}] to provide external consistency of the resulting
propensity-score-adjusted estimator to the known population mean $\overline {X}_{N}$.
To incorporate both population and sample-level information, similar to
Section~\ref{sec3}, the regression estimator $\hat{\theta}_{1,\mathrm{Reg}}$ of $\theta
_{1}$ can be written as
%
\begin{eqnarray}\label{A-1}
\hat{\theta}_{1,\mathrm{Reg}} &=& \hat{\theta}_{1}-
\widehat{B}_{\mathrm{Reg}}(\hat {\theta}_{x,1}-\overline {X}_{N})
\nonumber\\[-8pt]\\[-8pt]
\nonumber
&\cong& \hat{\theta}_{1}-B_{\mathrm{Reg}}(\hat{\theta}_{x,1}-
\overline {X}_{N}),
\end{eqnarray}
where $\hat{\theta}_{1}$ is defined in (\ref{1-5}),
\[
\hat{\theta}_{x,1}=N^{-1}\sum_{i \in A_{1}}
\omega_{i}\hat{\pi }^{-1}_{1i}x_{i},
\qquad\widehat{B}_{\mathrm{Reg}}=\sum_{i \in A_{1}}\omega
_{i}\hat{\pi}^{-1}_{1i}y_{1i}x'_{i}
\biggl(\sum_{i \in A_{1}}\omega _{i}\hat{
\pi}^{-1}_{1i}x_{i}x'_{i}
\biggr)^{-1}
\]
and $B_{\mathrm{Reg}}=p\lim\widehat{B}_{\mathrm{Reg}}$.
After ignoring the higher-order terms, it can be shown that
\begin{eqnarray}
\hat{\theta}_{1,\mathrm{Reg}} &=& \frac{1}{N}\sum
^{N}_{i=1} \biggl\{ \omega _{i}
\delta_{i}\frac{\delta_{1i}}{\pi
_{1i}}(y_{1i}-B_{\mathrm{Reg}}x_{i})-B_{1,\mathrm{Reg}}
\omega_{i}\delta_{i}\delta _{1i}\biggl(
\frac{\delta_{2i}}{\pi_{2i}}-1\biggr)h_{1i}
\nonumber
\\
&&\hspace*{111pt}{}- B_{2,\mathrm{Reg}}\biggl(\omega_{i}\delta_{i}
\frac{\delta_{1i}}{\pi
_{1i}}-1\biggr)+B_{\mathrm{Reg}}\overline {X}_{N} \biggr\},
\nonumber
\end{eqnarray}
where $B_{1,\mathrm{Reg}}=B_{1,y}-B_{\mathrm{Reg}}B_{1,x}$, $B_{2,\mathrm{Reg}}=B_{2,y}-B_{\mathrm{Reg}}B_{2,x}$,
\[
(B_{1,x}, B_{2,x}) = \sum_{i=1}^{N}(1-
\pi_{1i})x_{i}\bigl(0,x'_{i},y_{1i},1
\bigr)\pmatrix{
\displaystyle\sum_{i=1}^{N}\pi_{1i}(1-\pi_{2i})h_{1i}h'_{2i},0_{r \times1}
\vspace*{3pt}\cr
\displaystyle\sum_{i=1}^{N}(1-\pi_{1i}) \bigl(0,x'_{i},y_{1i},1\bigr)}^{-1}.
\]
Therefore, the asymptotic variance can be estimated as follows:
%
\begin{eqnarray} \label{evar1Reg}
\qquad \widehat{\operatorname{var}} ( \hat{\theta}_{1,\mathrm{Reg}} ) &=& \frac
{1}{N^{2}}
\sum_{i \in A_{1}}\omega_{i}\frac{1-\hat{\pi}_{1i}}{\hat
{\pi}^{2}_{1i}}(y_{1i}-
\widehat{B}_{\mathrm{Reg}}x_{i}-\widehat{B}_{2,\mathrm{Reg}})^{2}
\nonumber
\\
&&{}+ \frac{1}{N^{2}}\sum_{i \in A}\sum
_{j \in A}\widetilde{\Delta }_{ij}\frac{\hat{z}_{5i}}{\pi_{i}}
\frac{\hat{z}_{5j}}{\pi_{j}}
\nonumber\\[-8pt]\\[-8pt]
&&{} + \frac{1}{N^{2}}\sum_{i \in A_{2}}
\omega_{i}\frac{1-\hat{\pi
}_{2i}}{\hat{\pi}^{2}_{2i}}(\widehat{B}_{1,\mathrm{Reg}}h_{1i})^{2}\nonumber
\\
&&{}+ \frac{1}{N^{2}}\sum_{i \in A}\sum
_{j \in A}\widetilde{\Delta }_{ij}\frac{\hat{z}_{6i}}{\pi_{i}}
\frac{\hat{z}_{6j}}{\pi_{j}},\nonumber
\end{eqnarray}
where\vspace*{2pt} $\hat{z}_{5i}=\delta_{1i}\hat{\pi}^{-1}_{1i}(y_{1i}-\widehat{B}_{\mathrm{Reg}}x_{i}-\widehat{B}_{2,\mathrm{Reg}})$, $\hat{z}_{6i}=\delta_{1i}(\delta
_{2i}\hat{\pi}^{-1}_{2i}-1)\widehat{B}_{1,\mathrm{Reg}}h_{1i}$, $\widehat{B}_{1,\mathrm{Reg}}=\widehat{B}_{1,y}-\widehat{B}_{\mathrm{Reg}}\widehat{B}_{1,x}$, $\widehat{B}_{2,\mathrm{Reg}}=\widehat{B}_{2,y}-\widehat{B}_{\mathrm{Reg}}\widehat{B}_{2,x}$,
\begin{eqnarray*}
(\widehat{B}_{1,x}, \widehat{B}_{2,x}) &=& \sum
_{i \in A_{1}}\omega_{i}\hat{\pi}^{-1}_{1i}(1-
\hat{\pi }_{1i})x_{i}\bigl(0,x'_{i},y_{1i},1
\bigr)
\\
&&{}\times \pmatrix{ %
\displaystyle\sum_{i \in A_{2}}\omega_{i}\hat{\pi}^{-1}_{2i}(1-\hat{\pi}_{2i})h_{1i}h'_{2i}, 0_{r \times1}
\vspace*{3pt}\cr
\displaystyle\sum_{i \in A_{1}}\omega_{i}\hat{
\pi}^{-1}_{1i}(1-\hat{\pi }_{1i})
\bigl(0,x'_{i},y_{1i},1\bigr)}^{-1}.
\end{eqnarray*}

In addition, we can consider a design-optimal regression estimator, as
discussed in Remark~\ref{rem3.1}, that incorporates all information in an optimal
way. Specifically, the optimal regression estimator of $\theta_{1}$
can be written as
%
\begin{equation}
\hat{\theta}_{1,\mathrm{opt}}=\hat{\theta}_{1}-\widehat{B}_{\mathrm{opt}}(
\hat{\theta }_{x,1}-\overline {X}_{N}), \label{A-2}
\end{equation}
where $\widehat{B}_{\mathrm{opt}}=\widehat{\operatorname{cov}}( \hat{\theta}_{1},\hat
{\theta}_{x,1})\widehat{\operatorname{var}}^{-1}(\hat{\theta}_{x,1})$, with
\begin{eqnarray*}
\widehat{\operatorname{cov}}( \hat{\theta}_{1},\hat{\theta}_{x,1})
&=& \frac{1}{N^{2}}\sum_{i \in A_{1}}\omega_{i}
\frac{1-\hat{\pi
}_{1i}}{\hat{\pi}^{2}_{1i}}\hat{\eta}^{*}_{i}+\frac{1}{N^{2}}\sum
_{i \in A_{2}}\omega_{i}\frac{1-\hat{\pi}_{2i}}{\hat{\pi
}^{2}_{2i}}
\widehat{B}_{1,y}h_{1i}(\widehat{B}_{1,x}h_{1i})'
\\
&&{}+\frac{1}{N^{2}}\sum_{i \in A}\sum
_{j \in A}\widetilde{\Delta }_{ij}\frac{\hat{z}_{7i}}{\pi_{i}}
\frac{\hat{z}^{\prime}_{8j}}{\pi
_{j}},
\\
\widehat{\operatorname{var}}(\hat{\theta}_{x,1}) &=& \frac{1}{N^{2}}\sum
_{i\in A_{1}}\omega_{i}\frac{1-\hat{\pi}_{1i}}{\hat{\pi
}^{2}_{1i}}(x_{i}-
\widehat{B}_{2,x})^{\otimes2}
\\
&&{} +\frac{1}{N^{2}}\sum
_{i\in A_{2}}\omega_{i}\frac{1-\hat{\pi}_{2i}}{\hat{\pi
}^{2}_{2i}}(
\widehat{B}_{1,x}h_{1i})^{\otimes2}
\\
&&{}+ \frac{1}{N^{2}}\sum_{i \in A}\sum
_{j \in A}\widetilde{\Delta }_{ij}\frac{\hat{z}_{8i}}{\pi_{i}}
\frac{\hat{z}^{\prime}_{8j}}{\pi
_{j}},
\end{eqnarray*}
and
\begin{eqnarray*}
\hat{z}_{7i}&=&\delta_{1i}\hat{\pi}^{-1}_{1i}
\bigl\{ y_{1i}-\bigl(\delta _{2i}\hat{\pi}^{-1}_{2i}-1
\bigr)\hat{\pi}_{1i}\widehat{B}_{1,y}h_{1i}-\widehat{B}_{2,y} \bigr\},
\\
\hat{z}_{8i}&=&\delta_{1i}\hat{\pi}^{-1}_{1i}
\bigl\{ x_{i}-\bigl(\delta _{2i}\hat{\pi}^{-1}_{2i}-1
\bigr)\hat{\pi}_{1i}\widehat{B}_{1,x}h_{1i}-\widehat{B}_{2,x} \bigr\},
\end{eqnarray*}
$\hat{\eta}^{*}_{i}=x'_{i}y_{1i}-y_{1i}\widehat{B}'_{2,x}-\widehat{B}_{2,y}x'_{i}+\widehat{B}_{2,y}\widehat{B}'_{2,x}$. $\widehat{B}_{1,x}$, $\widehat{B}_{1,y}$, $\widehat{B}_{2,x}$ and $\widehat{B}_{2,y}$ are defined in Section~\ref{sec3}.

Now, the regression estimator of $\theta_{2}$ can be written as
\begin{eqnarray}
\hat{\theta}_{2,\mathrm{Reg}} &=& \hat{\theta}_{2}-\widehat{B}^{*}_{1,\mathrm{Reg}}(
\hat {h}_{2,1}-\hat{h}_{1,1} )-\widehat{B}^{*}_{2,\mathrm{Reg}}(
\hat{\theta }_{x,2}-\overline {X}_{N})
\nonumber
\\
&=& \hat{\theta}_{2}-B^{*}_{1,\mathrm{Reg}}(
\hat{h}_{2,1}-\hat{h}_{1,1} )-B^{*}_{2,\mathrm{Reg}}(
\hat{\theta}_{x,2}-\overline {X}_{N}),
\nonumber
\end{eqnarray}
where $\hat{\theta}_{x,2}=N^{-1}\sum_{i \in A_{2}}\omega_{i}\hat
{\pi}^{-1}_{1i}\hat{\pi}^{-1}_{2i}x_{i}$. $(\widehat{B}^{*}_{1,\mathrm{Reg}},\widehat{B}^{*}_{2,\mathrm{Reg}})$ is the regression coefficient.
It can be shown that
%
\begin{eqnarray}
\hat{\theta}_{2,\mathrm{Reg}} &=& \frac{1}{N}\sum
_{i=1}^{N}\biggl\{ \omega _{i}
\delta_{i}\frac{\delta_{1i}\delta_{2i}}{\pi_{1i}\pi
_{2i}}\bigl(y_{2i}-B^{*}_{2,\mathrm{Reg}}x_{i}
\bigr)-D^{*}_{1i,\mathrm{Reg}}\omega_{i}\delta _{i}
\delta_{1i}\biggl(\frac{\delta_{2i}}{\pi_{2i}}-1\biggr)h_{1i}\hspace*{-17pt}
\nonumber\\[-8pt]\\[-8pt]
\nonumber
&&\hspace*{127pt}{}- D^{*}_{2,\mathrm{Reg}}\biggl(\omega_{i}
\delta_{i}\frac{\delta_{1i}}{\pi
_{1i}}-1\biggr)+B^{*}_{2,\mathrm{Reg}}
\overline {X}_{N} \biggr\},\hspace*{-17pt}
\end{eqnarray}
where
\begin{eqnarray*}
D^{*}_{1i,\mathrm{Reg}} &=& \bigl\{ D_{1,y}+B^{*}_{1,\mathrm{Reg}}
\bigl(\pi ^{-1}_{1i}-D_{1,h_{1}}+B_{1,h_{1}}
\bigr)-B^{*}_{2,\mathrm{Reg}}D_{1,x} \bigr\},
\\
D^{*}_{2,\mathrm{Reg}} &=& \bigl\{ D_{2,y}-B^{*}_{1,\mathrm{Reg}}(D_{2,h_{1}}-B_{2,h_{1}})-B^{*}_{2,\mathrm{Reg}}D_{2,x}
\bigr\},
\end{eqnarray*}
and
\begin{eqnarray*}
(D_{1,x},D_{2,x}) &=& \sum^{N}_{i=1}x_{i}
\bigl\{ (1-\pi _{1i}) \bigl(0,x'_{i},y_{1i},1
\bigr)+(1-\pi_{2i}) \bigl(1,x'_{i},y_{2i},0
\bigr) \bigr\}
\\
&&{}\times \pmatrix{ %
\displaystyle\sum^{N}_{i=1}
\pi_{1i}(1-\pi_{2i})h_{1i}h'_{2i},
0_{r \times1}
\vspace*{3pt}\cr
\displaystyle\sum^{N}_{i=1}(1-
\pi_{1i}) \bigl(0,x'_{i},y_{1i},1
\bigr)}^{-1}.
\end{eqnarray*}

Also, similarly to (\ref{A-2}), the optimal regression estimator of
$\theta_{2}$ can be written as
%
\begin{equation}
\hat{\theta}_{2,\mathrm{opt}}=\hat{\theta}_{2}-\widehat{B}^{*}_{1,\mathrm{opt}}(
\hat {h}_{2,1}-\hat{h}_{1,1} )-\widehat{B}^{*}_{2,\mathrm{opt}}(
\hat{\theta }_{x,2}-\overline {X}_{N}), \label{opt2}
\end{equation}
where
\[
\bigl(\widehat{B}^{*}_{1,\mathrm{opt}},\widehat{B}^{*}_{2,\mathrm{opt}}
\bigr)=\widehat{\operatorname{cov}} \bigl\{ \hat{\theta}_{2}, (
\hat{h}_{2,1}-\hat{h}_{1,1},\hat{\theta }_{x,2}) \bigr
\} \bigl[\widehat{\operatorname{var}} \bigl\{ (\hat {h}_{2,1}-
\hat{h}_{1,1},\hat{\theta}_{x,2}) \bigr\} \bigr]^{-1}.
\]
$\widehat{\operatorname{cov}}(\hat{\theta}_{2},\hat{h}_{2,1}-\hat
{h}_{1,1})$, $\widehat{\operatorname{var}}(\hat{h}_{2,1}-\hat{h}_{1,1})$ are
defined in (\ref{Copt}),
\begin{eqnarray*}
\widehat{\operatorname{cov}}(\hat{h}_{2,1}-\hat{h}_{1,1},\hat{
\theta }_{x,2}) &\cong& \frac{1}{N^{2}}\sum
_{i \in A_{2}}\omega_{i}\frac
{1-\hat{\pi}_{2i}}{\hat{\pi}^{2}_{2i}}\biggl(
\frac{x_{i}}{\hat{\pi
}_{1i}}-\widehat{D}_{1,x}h_{1i}\biggr)\hat{
\eta}'_{i}
\\
&&{} -\frac{1}{N^{2}}\sum
_{i\in A_{1}}\omega_{i}\frac{1-\hat{\pi}_{1i}}{\hat{\pi
}^{2}_{1i}}(x_{2i}-
\widehat{D}_{2,x})\hat{\tau}'
\\
&&{}+ \frac{1}{N^{2}}\sum_{i \in A}\sum
_{j \in A}\widetilde{\Delta }_{ij}\frac{\hat{z}_{9i}}{\pi_{i}}
\frac{\hat{z}^{\prime}_{4j}}{\pi
_{j}},
\end{eqnarray*}
where
\[
\hat{z}_{9i}=\frac{\delta_{1i}}{\hat{\pi}_{1i}} \biggl\{\frac
{\delta_{2i}}{\hat{\pi}_{2i}}x_{i}-
\biggl(\frac{\delta_{2i}}{\hat{\pi
}_{2i}}-1\biggr)\hat{\pi}_{1i}\widehat{D}_{1,x}h_{1i}-
\widehat{D}_{2,x} \biggr\},
\]
%
$\widehat{\operatorname{cov}}(\hat{\theta}_{2},\hat{\theta}_{x,2})$ and
$\widehat{\operatorname{var}}(\hat{\theta}_{x,2})$ can be derived similarly.

\section{Simulation study}\label{sec5}
\subsection{Simulation one}\label{sec5.1}
In simulation one, we performed a limited simulation study to test the
performance of our proposed estimator and to perform a sensitivity
analysis of the model assumptions.
In the simulation study, we generated a finite population of size
$N={}$10,000. In the population, we generated $(x_{1i},x_{2i}, y_{1i},
y_{2i},z_{i})$, where
\begin{eqnarray*}
Z_{i} &=& 0.5+0.5X_{1i}+0.5X_{2i}+e_{i},
\\
Y_{1i} &=& Z_{i}+U_{1i}, \qquad Y_{2i}=Z_{i}+U_{2i},
\end{eqnarray*}
$X_{1i}, X_{2i}, U_{1i}, U_{2i}$ and $e_{i}$ are independently\vspace*{1pt} and
identically distributed with $N(1,1)$, $N(0,1)$, $N(0,\sigma
^{2}_{1})$, $N(0,\sigma^{2}_{2})$ with $\sigma_{1}=\sigma_{2}=0.59$
and $\exp(1)-1$, respectively. From the finite population, we
repeatedly generated two-phase samples with approximate sample sizes
$n_{1}=500$ and $n_{2}=300$ for the phase one and phase two samples,
respectively. We consider the following response mechanisms for the
first phase and second phase sampling indicators $\delta_{1i}$ and
$\delta_{2i}$:
\begin{longlist}[((M3))]
\item[(M1)] \textit{Linear Ignorable}
\[
\pi_{1i}=\frac{\exp(\phi_{0}+\phi_{1}X_{i})}{1+\exp(\phi
_{0}+\phi_{1}X_{i})},\qquad\pi_{2i}=
\frac{\exp(\phi^{*}_{0}+\phi
_{1}X_{i})}{1+\exp(\phi^{*}_{0}+\phi_{1}X_{i})},
\]
where $(\phi_{0},\phi_{1},\phi^{*}_{0})=(-3.2,0.3,0.2)$.

\item[(M2)] \textit{Linear Nonignorable Nonmeasurement error}
\begin{eqnarray*}
\pi_{1i}&=&\frac{\exp(\phi_{0}+\phi_{1}X_{i}+\phi
_{2}Y_{1i})}{1+\exp(\phi_{0}+\phi_{1}X_{i}+\phi_{2}Y_{1i})},
\\
\pi_{2i}&=&
\frac{\exp(\phi^{*}_{0}+\phi_{1}X_{i}+\phi
_{2}Y_{2i})}{1+\exp(\phi^{*}_{0}+\phi_{1}X_{i}+\phi_{2}Y_{2i})},
\end{eqnarray*}
where $(\phi_{0},\phi_{1},\phi_{2},\phi^{*}_{0})=(-3.4,0.3,0.1,0.5)$.

\item[(M3)] \textit{Complementary log--log Nonignorable Nonmeasurement error}
\begin{eqnarray*}
\pi_{1i}&=&1-\exp \bigl\{ -\exp(\phi_{0}+
\phi_{1}X_{i}+\phi _{2}Y_{1i}) \bigr\},
\\
\pi_{2i}&=& 1-\exp \bigl\{ -\exp\bigl(\phi^{*}_{0}+
\phi_{1}X_{i}+\phi _{2}Y_{2i}\bigr)
\bigr\},
\end{eqnarray*}
where $(\phi_{0},\phi_{1},\phi_{2},\phi^{*}_{0})=(-3,0.1,-0.1,-0.1)$.

\item[(M4)] \textit{Probit Nonignorable Nonmeasurement error}
\[
\pi_{1i}=\Phi(\phi_{0}+\phi_{1}X_{i}+
\phi_{2}Y_{1i}),\qquad\pi _{2i}=\Phi\bigl(
\phi^{*}_{0}+\phi_{1}X_{i}+
\phi_{2}Y_{2i}\bigr),
\]
where $(\phi_{0},\phi_{1},\phi_{2},\phi^{*}_{0})=(-2,0.2,0.2,0.1)$.

\item[(M5)] \textit{Linear Nonignorable Measurement error}
\begin{eqnarray*}
\pi_{1i}&=&\frac{\exp(\phi_{0}+\phi_{1}X_{i}+\phi_{2}Z_{i})}{1+\exp
(\phi_{0}+\phi_{1}X_{i}+\phi_{2}Z_{i})},
\\
\pi_{2i}&=&
\frac{\exp
(\phi^{*}_{0}+\phi_{1}X_{i}+\phi_{2}Z_{i})}{1+\exp(\phi
^{*}_{0}+\phi_{1}X_{i}+\phi_{2}Z_{i})},
\end{eqnarray*}
where $(\phi_{0},\phi_{1},\phi_{2},\phi^{*}_{0})=(-3.4,0.3,0.1,0.5)$.

\item[(M6)] \textit{Complementary log--log Nonignorable Measurement error}
\begin{eqnarray*}
\pi_{1i}&=&1-\exp \bigl\{ -\exp(\phi_{0}+
\phi_{1}X_{i}+\phi _{2}Z_{i}) \bigr\},
\\
\pi_{2i}&=& 1-\exp \bigl\{ -\exp\bigl(\phi^{*}_{0}+
\phi_{1}X_{i}+\phi _{2}Z_{i}\bigr)
\bigr\},
\end{eqnarray*}
where $(\phi_{0},\phi_{1},\phi_{2},\phi^{*}_{0})=(-3,0.1,-0.1,-0.1)$.

\item[(M7)] \textit{Probit Nonignorable Measurement error}
\[
\pi_{1i}=\Phi(\phi_{0}+\phi_{1}X_{i}+
\phi_{2}Z_{i}),\qquad\pi _{2i}=\Phi\bigl(
\phi^{*}_{0}+\phi_{1}X_{i}+
\phi_{2}Z_{i}\bigr),
\]
where $(\phi_{0},\phi_{1},\phi_{2},\phi^{*}_{0})=(-2,0.2,0.2,0.1)$.
\end{longlist}
We used $B=2000$ Monte Carlo sample size. The ``working'' model for
the propensity score estimation is the Linear Nonignorable model (M2).
From each sample, we computed the following four estimators for $\theta
_{1}=E(Y_1)$:
\begin{longlist}[(3)]
\item[(1)] Naive: Calibration estimator which assumes ignorable missing mechanism;
\item[(2)] PS: Proposed propensity score estimator, as defined in (\ref{1-5});
\item[(3)] REG: Proposed regression estimator, as defined in (\ref{A-1}); and
\item[(4)] OPT: Proposed optimal estimator, as defined in (\ref{A-2}).
\end{longlist}
We also computed the following five estimators for $\theta_{2}=E(Y_2)$:
\begin{longlist}[(1)]
\item[(1)] Naive: Calibration estimator which assumes ignorable missing mechanism;
\item[(2)] PS: Proposed propensity score estimator, as defined in (\ref{1-6});
\item[(3)] REG: Proposed regression estimator by incorporating first-phase
sample information, as defined in (\ref{reg});
\item[(4)] OPT1: Proposed optimal estimator by incorporating first-phase
sample information, as defined in Remark~\ref{rem3.1}; and
\item[(5)] OPT2: Proposed optimal estimator by incorporating both
first-phase sample and population information, as defined in (\ref{opt2}).
\end{longlist}

%
\begin{table}[t]
\tabcolsep=0pt
\caption{Simulation results of the point estimators and variance
estimators under models \textup{(M1)--(M4)} for simulation one}\label{tab1}
\begin{tabular*}{\tablewidth}{@{\extracolsep{\fill}}@{}lccd{2.4}ccd{2.4}@{}}
\hline
\textbf{Model} & \textbf{Parameter} & \textbf{Method} &
\multicolumn{1}{c}{\textbf{Bias}} & \multicolumn{1}{c}{\textbf{SE}} & \multicolumn{1}{c}{\textbf{RMSE}} & \multicolumn{1}{c@{}}{\textbf{RB}}\\
\hline
(M1)& $\theta_{1}$ & Naive & -0.0029 & 0.0751 & 0.0751 &\multicolumn{1}{c@{}}{N/A} \\
& & PS & 0.0071 & 0.1647 & 0.1648 & 0.0363 \\
& & REG & 0.0108 & 0.1622 & 0.1625 & 0.0160 \\
& & OPT & 0.0077 & 0.1556 & 0.1557 & 0.0543
\\[3pt]
& $\theta_{2}$ & Naive & -0.0034 & 0.0907 &0.0908 & \multicolumn{1}{c@{}}{N/A} \\
& & PS & 0.0069 & 0.1559 & 0.1560 & 0.0342 \\
& & REG & 0.0072 & 0.1584 & 0.1586 & 0.0123 \\
& & OPT1 & 0.0074 & 0.1582 & 0.1584 & -0.0153 \\
& & OPT2 & 0.0082 & 0.1496 & 0.1498 & -0.0182
\\[6pt]
(M2) & $\theta_{1}$ & Naive & 0.1662 & 0.0838 & 0.1861 &\multicolumn{1}{c@{}}{N/A} \\
& & PS & -0.0026 & 0.1786 & 0.1786 & -0.0093 \\
& & REG & -0.0013 & 0.1708 & 0.1709 & 0.0034 \\
& & OPT & -0.0001 & 0.1670 & 0.1670 & 0.0118
\\[3pt]
& $\theta_{2}$ & Naive & 0.1857 & 0.0954 &0.2088 & \multicolumn{1}{c@{}}{N/A} \\
& & PS & 0.0007 & 0.1645 & 0.1644 & -0.0217 \\
& & REG & 0.0002 & 0.1656 & 0.1656 & -0.0160 \\
& & OPT1 & 0.0022 & 0.1633 & 0.1633 & -0.0294 \\
& & OPT2 & 0.0030 & 0.1524 & 0.1524 & -0.0238
\\[6pt]
(M3) & $\theta_{1}$ & Naive & -0.1578 & 0.0721 & 0.1735 &\multicolumn{1}{c@{}}{N/A} \\
& & PS& 0.0712 & 0.1957 & 0.2083 & -0.0077 \\
& & REG & 0.0910 & 0.1963 & 0.2164 & -0.0316 \\
& & OPT & 0.0777 & 0.1846 & 0.2003 & 0.0283
\\[3pt]
& $\theta_{2}$ & Naive & -0.2137 & 0.0843 &0.2297 & \multicolumn{1}{c@{}}{N/A} \\
& & PS & 0.0490 & 0.1795 & 0.1860 & 0.0215 \\
& & REG & 0.0546 & 0.1889 & 0.1966 & -0.0849 \\
& & OPT1 & 0.0420 & 0.1833 & 0.1880 & -0.0506 \\
& & OPT2 & 0.0523 & 0.1759 & 0.1835 & -0.0506
\\[6pt]
(M4) & $\theta_{1}$ & Naive & -0.2593 & 0.0671 & 0.2678 &\multicolumn{1}{c@{}}{N/A} \\
& & PS& -0.0325 & 0.1822 & 0.1850 & 0.0036 \\
& & REG & -0.0830 & 0.1714 & 0.1904 & 0.0833 \\
& & OPT & -0.0707 & 0.1701 & 0.1842 & 0.0266
\\[3pt]
& $\theta_{2}$ & Naive & -0.2689 & 0.0765 &0.2795 & \multicolumn{1}{c@{}}{N/A} \\
& & PS & -0.0244 & 0.1544 & 0.1563 & 0.0823 \\
& & REG & -0.0180 & 0.1678 & 0.1688 & -0.0914 \\
& & OPT1 & -0.0242 & 0.1630 & 0.1648 & -0.0549 \\
& & OPT2 & -0.0607 & 0.1583 & 0.1695 & -0.0945 \\
\hline
\end{tabular*}\vspace*{-3pt}
\end{table}

%
\begin{table}
\tabcolsep=0pt
\caption{Simulation results of the point estimators and variance
estimators under models \textup{(M5)--(M7)} for simulation one}\label{tab2}
\begin{tabular*}{\tablewidth}{@{\extracolsep{\fill}}@{}lccd{2.4}ccd{2.4}@{}}
\hline
\textbf{Model} & \textbf{Parameter} & \textbf{Method} &
\multicolumn{1}{c}{\textbf{Bias}} & \multicolumn{1}{c}{\textbf{SE}} & \multicolumn{1}{c}{\textbf{RMSE}} & \multicolumn{1}{c@{}}{\textbf{RB}}\\
\hline
(M5)& $\theta_{1}$ & Naive & 0.1242 & 0.0794 & 0.1474 & \multicolumn{1}{c@{}}{N/A} \\
& & PS & -0.0312 & 0.1788 & 0.1814 & -0.0147 \\
& & REG & -0.0278 & 0.1707 & 0.1729 & -0.0236 \\
& & OPT & -0.0270 & 0.1668 & 0.1690 & -0.0055
\\[3pt]
& $\theta_{2}$ & Naive & 0.1649 & 0.0911 &0.1884 & \multicolumn{1}{c@{}}{N/A} \\
& & PS & -0.0037 & 0.1620 & 0.1620 & 0.0001 \\
& & REG & -0.0049 & 0.1648 & 0.1648 & -0.0150 \\
& & OPT1 & -0.0014 & 0.1632 & 0.1632 & -0.0364 \\
& & OPT2 & 0.0007 & 0.1525 & 0.1525 & -0.0412
\\[6pt]
(M6)& $\theta_{1}$ & Naive & -0.1269 & 0.0725 & 0.1462 &\multicolumn{1}{c@{}}{N/A} \\
& & PS & 0.1129 & 0.1993 & 0.2290 & 0.0032 \\
& & REG & 0.1309 & 0.2006 & 0.2396 & -0.0517 \\
& & OPT & 0.1196 & 0.1914 & 0.2256 & -0.0006
\\[3pt]
& $\theta_{2}$ & Naive & -0.1926 & 0.0838 &0.2100 & \multicolumn{1}{c@{}}{N/A} \\
& & PS & 0.0787 & 0.1863 & 0.2023 & 0.0008 \\
& & REG & 0.0831 & 0.1918 & 0.2090 & -0.0624 \\
& & OPT1 & 0.0757 & 0.1875 & 0.2023 & -0.0437 \\
& & OPT2 & 0.0843 & 0.1797 & 0.1985 & -0.0464
\\[6pt]
(M7)& $\theta_{1}$ & Naive & -0.1922 & 0.0660 & 0.2032 &\multicolumn{1}{c@{}}{N/A} \\
& & PS& 0.0356 & 0.1705 & 0.1742 & 0.0287 \\
& & REG & -0.0174 & 0.1635 & 0.1644 & 0.0847 \\
& & OPT & -0.0002 & 0.1664 & 0.1664 & 0.0006
\\[3pt]
& $\theta_{2}$ & Naive & -0.2526 & 0.0771 &0.2641 & \multicolumn{1}{c@{}}{N/A} \\
& & PS & 0.0005 & 0.1626 & 0.1626 & 0.0240 \\
& & REG & 0.0017 & 0.1626 & 0.1626 & 0.0128 \\
& & OPT1 & -0.0017 & 0.1630 & 0.1630 & -0.0138 \\
& & OPT2 & -0.0375 & 0.1579 & 0.1623 & -0.0399 \\
\hline
\end{tabular*}
\end{table}

\begin{figure}

\includegraphics{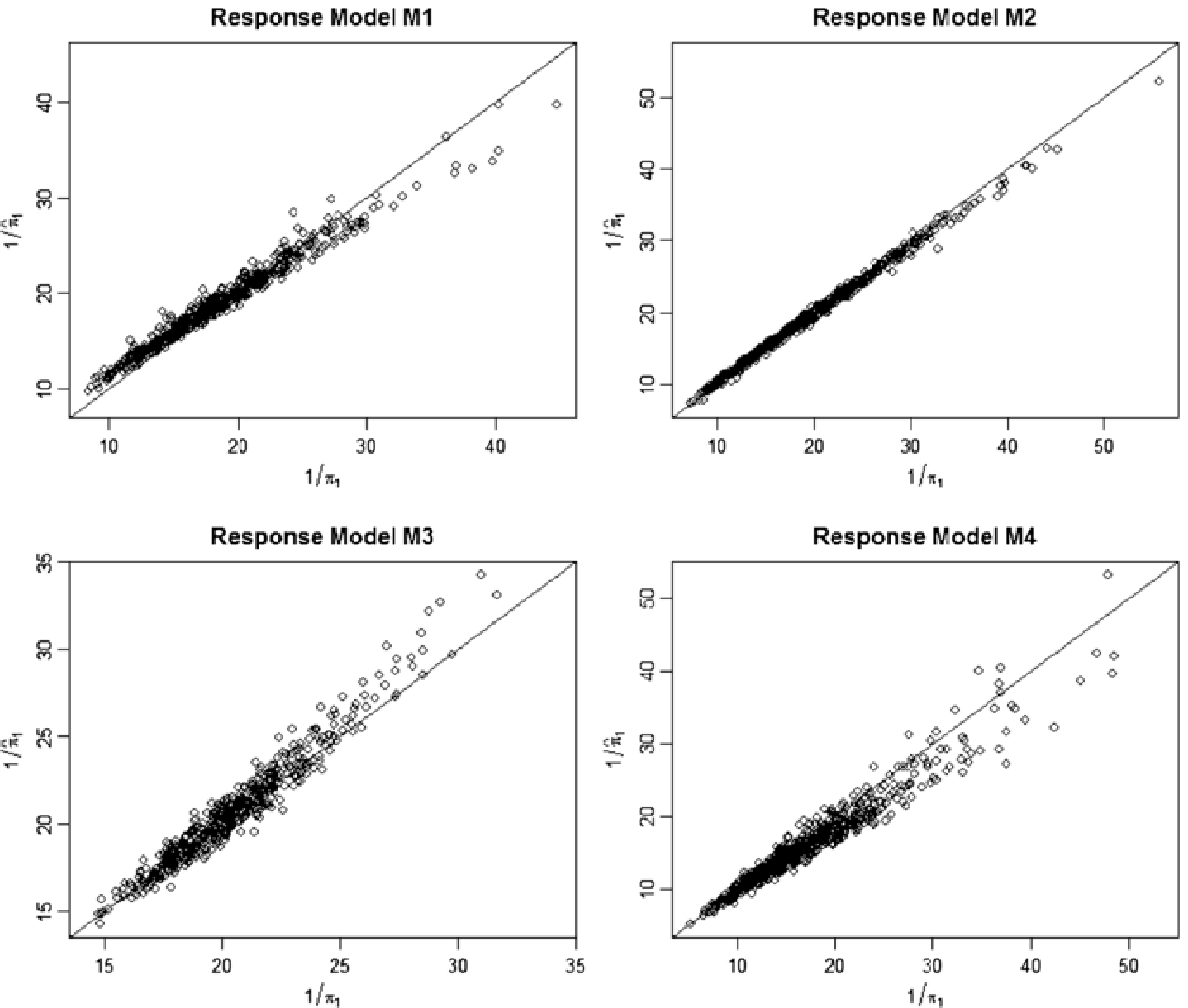}

\caption{Plots of the fitted values on the true value of the inverse
response probability under models~\textup{(M1)--(M4)}.}\label{fig1}
\end{figure}

The simulation results for point and variance estimations are given in
Tables~\ref{tab1} for models (M1)--(M4) and~\ref{tab2} for models (M5)--(M7).
Specifically, we calculated the Biases, Standard Errors (SE), Root
Mean Squared Errors (RMSE) of the point estimators and Relative Bias
(RB) of the proposed variance estimators. Under models (M1) and (M2),
the proposed estimators show negligible biases, which confirms our
theory. Furthermore, the proposed estimators show modest biases even
under models (M3)--(M7) where the assumed response model is not equal to
the true response model. To investigate the effect of using incorrect
response models, we made plots of $1/\hat{\pi}_i$ on $1/\pi_i$ under
four different response models \mbox{(M1)--(M4)}, where $\hat{\pi}_i$ are the
fitted values of the response probabilities under the working model and
$\pi_i$ are the true response probabilities. In Figure~\ref{fig1},
the plot under model (M3) shows that most of the sample observations
are located above the line of $y=x$, which
suggests that many observations are assigned to bigger propensity
weights $(1/\hat{\pi}_i)$ than necessary. Thus, the resulting PS
estimator can be positively biased, as confirmed in Table~\ref{tab1}. On the
other hand, the plot under model~(M4) shows the opposite phenomenon and
the resulting estimator is negatively biased. According to the plots,
we find that the fitted values are highly correlated with the true
values even under the wrong models. Thus, the bias of the PS estimators
is also modest under (M3)--(M4). Similar results have been found for
(M5)--(M7). The regression estimator and the optimal estimator are more
efficient than the PS estimator, and the optimal estimator has the
smallest variance, which is consistent with the theory.
The PS estimator of $\theta_2$ is more efficient than that of $\theta
_1$ since $y_{2i}$ is well approximated by $y_{1i}$, which is a part of
$h_{1i}$ when computing the model parameters.
Variance estimators show negligible relative biases in most cases. The
simulation study suggests that the proposed estimator is robust against
the failure of the assumed response models.

%

\subsection{Simulation two}\label{sec5.2}
In this simulation study, we consider only model (M5) Linear
Nonignorable Measurement error. In order to test the sensitivity of
model misspecification of equal coefficients for two-phase response
mechanisms, we perform simulation studies by using the following
response mechanisms:
\[
\pi_{1i}=\frac{\exp(\phi_{0}+\phi_{1}x_{1i}+\phi
_{2}z_{i})}{1+\exp(\phi_{0}+\phi_{1}x_{1i}+\phi_{2}z_{i})},\qquad \pi_{2i}=
\frac{\exp(\phi^{*}_{0}+\phi_{1}x_{1i}+\phi
^{*}_{2}z_{i})}{1+\exp(\phi^{*}_{0}+\phi_{1}x_{1i}+\phi^{*}_{2}z_{i})},
\]
where $(\phi_{0},\phi_{1},\phi_{2},\phi^{*}_{0},\phi
^{*}_{2})=(-3.4,0.3,0.1+d,0.5,0.1)$. Specifically, we consider the
following three cases:
\begin{longlist}[(C3)]
\item[(C1)] Small Difference: $d=0.01$.
\item[(C2)] Medium Difference: $d=0.05$.
\item[(C3)] Big Difference: $d=0.1$.
\end{longlist}
Other setups of simulation two are the same as that for simulation one.
The results are presented in Table~\ref{tab3}. According to Table~\ref{tab3}, the biases
for all the proposed estimators increase when the discrepancy parameter
$d$ increases, but the biases are still very modest. The variance
estimators show very small relative biases, which confirms the
stability of the variance estimators. Other patterns of the results are
similar to that for simulation one.

%
\begin{table}
\caption{Simulation results of the point estimators and variance
estimators under model \textup{(M5)} for simulation two}\label{tab3}
\begin{tabular*}{\tablewidth}{@{\extracolsep{\fill}}@{}lccd{2.4}ccd{2.4}@{}}
\hline
\textbf{Model} & \textbf{Parameter} & \textbf{Method} &
\multicolumn{1}{c}{\textbf{Bias}} & \multicolumn{1}{c}{\textbf{SE}} & \multicolumn{1}{c}{\textbf{RMSE}} & \multicolumn{1}{c@{}}{\textbf{RB}}\\
\hline
(C1)& {$\theta_{1}$} & Naive & 0.1426 & 0.0837 & 0.1653 &\multicolumn{1}{c@{}}{N/A} \\
& & PS & -0.0199 & 0.1731 & 0.1742 & 0.0228 \\
& & REG & -0.0213 & 0.1627 & 0.1641 & 0.0279 \\
& & OPT & -0.0194 & 0.1607 & 0.1618 & 0.0281
\\[3pt]
& \multirow{5}{*}{$\theta_{2}$} & Naive & 0.1852 & 0.0929 &0.2071 & \multicolumn{1}{c@{}}{N/A} \\
& & PS & 0.0066 & 0.1607 & 0.1608 & 0.0080 \\
& & REG & 0.0059 & 0.1627 & 0.1628 & 0.0026 \\
& & OPT1 & 0.0076 & 0.1605 & 0.1607 & -0.0099 \\
& & OPT2 & 0.0064 & 0.1486 & 0.1487 & -0.0071
\\[6pt]
(C2)& {$\theta_{1}$} & Naive & 0.2043 & 0.0802 & 0.2194 &\multicolumn{1}{c@{}}{N/A} \\
& & PS & 0.0384 & 0.1726 & 0.1768 & -0.0213 \\
& & REG & 0.0239 & 0.1609 & 0.1626 & 0.0157 \\
& & OPT & 0.0295 & 0.1601 & 0.1628 & -0.0048
\\[3pt]
& \multirow{5}{*}{$\theta_{2}$} & Naive & 0.2457 & 0.0892 &0.2614 & \multicolumn{1}{c@{}}{N/A} \\
& & PS & 0.0610 & 0.1597 & 0.1709 & -0.0182 \\
& & REG & 0.0599 & 0.1611 & 0.1718 & -0.0168 \\
& & OPT1 & 0.0628 & 0.1602 & 0.1720 & -0.0427 \\
& & OPT2 & 0.0529 & 0.1480 & 0.1572 & -0.0273
\\[6pt]
(C3)& {$\theta_{1}$} & Naive & 0.2992 & 0.0797 & 0.3096 &\multicolumn{1}{c@{}}{N/A} \\
& & PS & 0.1093 & 0.1678 & 0.2002 & -0.0221 \\
& & REG & 0.0763 & 0.1605 & 0.1777 & -0.0131 \\
& & OPT & 0.0887 & 0.1597 & 0.1826 & -0.0446
\\[3pt]
& $\theta_{2}$ & Naive & 0.3533 & 0.0931 &0.3654 & \multicolumn{1}{c@{}}{N/A} \\
& & PS & 0.1357 & 0.1566 & 0.2072 & -0.0078 \\
& & REG & 0.1353 & 0.1587 & 0.2085 & -0.0115 \\
& & OPT1 & 0.1374 & 0.1573 & 0.2089 & -0.0355 \\
& & OPT2 & 0.1158 & 0.1498 & 0.1893 & -0.0644 \\
\hline
\end{tabular*}
\end{table}

\section{Empirical study}\label{sec6}

The proposed two-phase propensity score estimator is applied to the
data obtained from the 2012 Iowa Caucus Survey (ICS). The Iowa
political party caucuses are a significant component of the
presidential candidate selection process.
In 2011, two caucus polls were conducted to be implemented prior to the
January 2012 Iowa Republican Caucus. In the first poll, approximately
1200 registered Republicans and Independents (no party) were interviewed
in November 2011. The second poll was a follow-up conducted in December
2011 with the November 2011 respondents to identify changes in their
voting preferences.

The sampling frame for the November 2011 poll was constructed from the
Iowa voter registry provided by the Iowa Secretary of State. The
telephone numbers on the list were reported by voters at the time of
their registration and therefore included both landlines and cell phone
numbers. A stratified systematic sampling design was used to select the
initial sample. Five variables were used to create strata or sorting
variables to ensure spread across the range of variation in age, voter
activity, geography, gender and party affiliation. One indicator
variable was created to differentiate voters 35 years or above from
younger voters, and a second indicator variable defined whether a voter
had attended one or more of the previous five primaries. Three
additional variables used in designing the sample were congressional
district, registered party and gender.

Strata were defined by party affiliation, congressional district, the
age indicator and the prior primary attendance indicator. Within
parties, sample size allocation incorporated an oversampling of primary
attendees to maximize the chances of reaching likely caucus attendees.
Sample allocation across the remaining strata was defined in proportion
to the number of voters in each stratum. The stratified design was
implemented using a systematic probability proportional to size
selection scheme. The size measure was based on the relative proportion
of voters in each stratum. For each party list, the systematic
selection scheme was applied to a list of voters sorted by
congressional district, age indicator, previous primary attendance
indicator and gender.

\begin{figure}[b]

\includegraphics{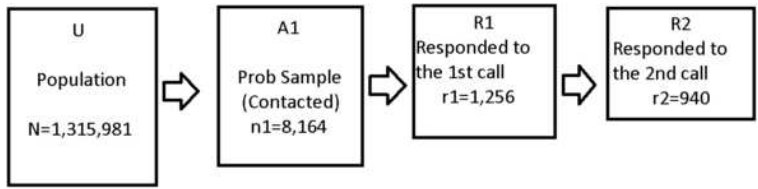}

\caption{Sample structure of 2012 Iowa Caucus Survey.}\label{fig2}
\end{figure}

A sample of 9000 voters was selected for the November 2011 poll,
consisting of 6000 Republicans and 3000 Independents. Telephone
numbers were unavailable for 836 of the sampled voters. The remaining
8164 sample households were contacted. Excluding 190 noneligible
numbers, 1256 registered voters were finally interviewed from the
November poll, which resulted in a 15.8 percent response rate.
The November survey of registered Republicans and Independents
contained questions related to anticipated caucus attendance,
candidates of choice and opinions on candidate characteristics, as well
as demographic and background items.
In the December 2011 poll, 1256 respondents from the November poll
were contacted again for a follow-up survey and 940 interviews were
completed, leading to a 74.9 percent response rate. Figure~\ref{fig2} summarizes
the two-phase sampling structure of the 2012 Iowa Caucus Survey.

%

To apply our proposed method to the ICS data, let $Y$ be the reported
value of the ``First Choice'' candidate. After preliminary analysis, we
decided to use $X=(\mathrm{Party},\mathrm{Age})$ as the auxiliary variable in the
propensity model. The auxiliary variable has a known total at the
population level and is also related to the survey participation rate.
The population size is $N={}$1,315,981. Denote $DY_{1}$ and $DY_{2}$ as
the dummy variables of ``First Choice'' based on the first sample
$A_{1}$ and the second sample $A_{2}$, and let $DX$ be the dummy
variable based on $X$. Then the parameters of interest are
\[
\theta_{1}=\frac{\sum_{i \in U}Z_{i}DY_{1i}}{\sum_{i \in U}Z_{i}}
\]
and
\[
\theta_{2}=\frac{\sum_{i \in U}Z_{i}DY_{2i}}{\sum_{i \in U}Z_{i}},
\]
where $Z_{i}$ is the indicator of ``Caucus Attendance'' for unit $i$.
That is, $Z_i=1$ if ``Caucus Attendance${}={}$Definitely attend'' or
``Caucus Attendance${}={}$Likely to attend.'' The outcome of the Iowa
Caucus on January 3, 2012 is
%
\begin{equation}
\theta_{0}=(24.5 \%, 10.3\%, 21.4\%,
43.7\%) \label{true}
\end{equation}
for ``First Choice'' candidate: Romney, Perry, Paul, Others.
Note that our parameters $\theta_1$ and $\theta_2$ are not
necessarily equal to $\theta_0$, although they may be close for
certain candidates.

The propensity model used for the proposed estimator is
%
\begin{equation}
\pi_{1i}(\phi)=\frac{\exp(\phi_{0}+\phi'_{1}DX_{i}+\phi
'_{2}DY_{1i})}{1+\exp(\phi_{0}+\phi'_{1}DX_{i}+\phi'_{2}DY_{1i})} \label{m1}
\end{equation}
and
%
\begin{equation}
\pi_{2i}\bigl(\phi^{*}\bigr)=\frac{\exp(\phi^{*}_{0}+\phi'_{1}DX_{i}+\phi
'_{2}DY_{2i})}{1+\exp(\phi^{*}_{0}+\phi'_{1}DX_{i}+\phi'_{2}DY_{2i})},
\label{m2}
\end{equation}
where $DX_{i}=(DX_{1i},DX_{2i})'$,
$DY_{1i}=(DY_{11i},DY_{12i},DY_{13i},DY_{14i},DY_{15i})'$ and
$DY_{2i}=(DY_{21i},DY_{22i},DY_{23i},DY_{24i},DY_{25i})'$.
Using the proposed methods in Section~\ref{sec3}, we obtain parameter estimates
for the selection model. The estimated parameters are given in Table~\ref{tab4}.
Table~\ref{tab4} shows that variables $DX_{1}$, $DX_{2}$ and $DY_{11}$ have
significant effects on the selection mechanisms, which supports our
model for nonignorable sample selection.

%
\begin{table}[b]
\tabcolsep=0pt
\caption{Estimated coefficients in the propensity model}\label{tab4} 
\begin{tabular*}{\tablewidth}{@{\extracolsep{\fill}}@{}lcccccc@{}} 
\hline
\textbf{Coefficient} & \textbf{Age} & \textbf{Party} & \textbf{Romney} & \textbf{Perry} & \textbf{Paul} & \textbf{Others}\\
\hline
Est & 0.588 & 0.782 & 0.991 & 0.454 & 0.866 & 1.307 \\
SE & 0.266 & 0.251 & 0.454 & 0.663 & 0.841 & 0.985 \\
$t$-value & 2.211 & 3.116 & 2.183 & 0.685 & 1.030 &1.327 \\
\hline
\end{tabular*}
\end{table}

We consider three estimators for estimating $\theta_t$ for $t=1,2$:
(i) the naive estimator (Naive) based on the respondents, computed by
$\hat{\theta}_{tN}=\sum_{i \in A_{t}}Z_{i}DY_{ti}/\break\grave{}(\sum_{i \in
A_{t}}Z_{i})$; (ii) the ignorable-response estimator (Ignorable),
computed by
$\hat{\theta}_{tIE}=\sum_{i \in A_{t}}\omega_{ti}Z_{i}DY_{ti}/ (\sum_{i \in A_{t}}\omega_{ti}Z_{i})$,
where $\omega_{ti}$ is the propensity score obtained by assuming the
ignorable adjustment weight which is obtained by setting $\phi_2=0$ in
the sample selection model; and (iii) the proposed propensity score
estimator using nonignorable sample selection models in (\ref{m1}) and
(\ref{m2}).
The proposed propensity score estimators are computed by (\ref{1-5})
and (\ref{1-6}).

%
\begin{table}
\tabcolsep=0pt
\caption{Estimated parameters (s.e.) for 2012 Iowa Caucus Survey results}\label{tab5} 
\begin{tabular*}{\tablewidth}{@{\extracolsep{\fill}}@{}lccccc@{}}
\hline
\textbf{Survey} & \textbf{Method} & \textbf{Romney} & \textbf{Perry} & \textbf{Paul} & \textbf{Others}\\
\hline
November & Naive & 0.340 & 0.108 & 0.130 & 0.422 \\
& Ignorable & 0.316 & 0.103 & 0.146 & 0.435 \\
& Proposed & 0.303 & 0.106 & 0.093 & 0.499 \\
& & (0.062) & (0.039) & (0.107) & (0.046)
\\[3pt]
December & Naive & 0.281 & 0.140 & 0.131 & 0.448 \\
& Ignorable & 0.270 & 0.144 & 0.148 & 0.437 \\
& Proposed &0.244 & 0.134 & 0.112 & 0.509 \\
& & (0.043) & (0.026) & (0.046) & (0.036) \\
\hline
\end{tabular*}
\tabnotetext[]{}{Standard errors are in parentheses.}
\end{table}

The results for point estimation are given in Table~\ref{tab5}. The proposed
estimates are closer to the Iowa Caucus results in (\ref{true}) than
the other estimates for Romney and Perry. Furthermore, the proposed
method enables us to compute the estimated standard errors of the point
estimates using the theory discussed in Section~\ref{sec3}. The estimated
standard errors in the December 2011 survey estimates are smaller than
those in the November 2011 survey estimates, which is consistent with
our findings in Section~\ref{sec5}. However, the estimates for Paul and Others
are further away from the reported true values compared to other
estimators, which are not that encouraging. It may be due to the
uncontrolled time effect.

\section{Concluding remarks}\label{sec7}

Estimators from self-selected samples can suffer from selection bias.
Propensity score weighting using demographic variables can reduce
selection bias, but the bias may remain important if survey
participation depends on the study variable itself. We make assumptions
about the selection mechanism that explicitly include the study
variable in the selection model. To estimate the model parameters, we
propose obtaining a second survey from the original self-selected
sample. If the second survey has questions similar to the first one, we
may assume that the regression coefficients for the explanatory
variables in the propensity model are the same as for the original
sample. The propensity model is then identified and the model
parameters can be estimated using a generalized method of moments. The
proposed method also permits estimation of the standard errors of the
estimated parameters.

As mentioned in Remark \ref{rem3.2}, our proposed approach is equivalent to the
measurement error model approach of \citet{r19} for propensity score weighting. Also, two limited simulation
studies in Section~\ref{sec5} suggest that the proposed method is robust against
the failure of the assumed response models and individual-level
heterogeneity in the propensity to respond that persists over time. The
proposed estimators have modest biases even when the equality of
coefficients for the two response mechanisms violate in a certain range.

The proposed method provides a useful tool for analyzing voluntary
samples as well as nonignorable nonresponse problems for survey data
and, in particular, web-based panel surveys. In a panel survey, the
same sample can be contacted several times and the proposed two-phase
estimation approach can be extended to multiphase estimation. This is a
topic of future study.

\section*{Acknowldegments}
We thank two anonymous
referees and the Associate Editor for their constructive comments which have helped
to improve the quality of the paper.
The data collection process was prepared and conducted by staff from
the Center for Survey Statistics and Methodology (CSSM) and Survey and
Behavioral Research Services (SBRS) at Iowa State University.

\begin{supplement}[id=suppA]
\sname{Supplement Derivations}
\stitle{Supplement Derivations for Sections~\ref{sec3} and \ref{sec4}}
\slink[doi]{10.1214/14-AOAS746SUPPA} 
\sdatatype{.pdf}
\sfilename{aoas746\_suppA.pdf}
\sdescription{Details of derivations for variances and variance
estimators in Sections~\ref{sec3}~and~\ref{sec4}.}
\end{supplement}

\begin{supplement} [id=suppB]
\sname{Supplement Justifications}
\stitle{Supplement Justifications for Remark \ref{rem3.2}}
\slink[doi]{10.1214/14-AOAS746SUPPB} 
\sdatatype{.pdf}
\sfilename{aoas746\_suppB.pdf}
\sdescription{Details of justifications for Remark \ref{rem3.2}.}
\end{supplement}


\printaddresses
\end{document}